\documentclass{emulateapj}
\shortauthors{Xing et al.}

\newcommand{\msp}{2FGL~J0523.3$-$2530}

\newcommand{\fermi}{\textit{Fermi}}
\newcommand{\gr}{$\gamma$-ray}

\begin{document}

\title{\textit{Fermi} Variability Study of the Candidate Pulsar Binary 2FGL~J0523.3$-$2530}

\author{Yi Xing\altaffilmark{1}, 
Zhongxiang Wang\altaffilmark{1}, 
C.-Y. Ng\altaffilmark{2}
}

\altaffiltext{1}{Shanghai Astronomical Observatory, Chinese Academy of Sciences,
80 Nandan Road, Shanghai 200030, China}

\altaffiltext{2}{Department of Physics, The University of Hong Kong,
Pokfulam Road, Hong Kong}

\begin{abstract}
The \textit{Fermi} source 2FGL~J0523.3$-$2530 has recently been identified
as a candidate millisecond pulsar binary with an orbital period of
16.5 hrs. We have carried out detailed studies of the source's emission 
properties by analyzing data taken with the \textit{Fermi} Large Area 
Telescope in the 0.2--300 GeV energy range. Long-term, yearly variability
from the source has been found, with a factor of 4 flux variations in 
1--300 GeV.  From spectral analysis, we find an extra spectral 
component at 2--3 GeV that causes the source brightening. 
While no orbital modulations have been found from 
the \textit{Fermi} data over the whole period of 2008--2014,
orbital modulation in the source's $>$2 GeV emission is detected
during the last 1.5 yrs of the \textit{Fermi} observation.
Our results support the millisecond pulsar binary nature 
of 2FGL~J0523.3$-$2530. Multi-wavelength observations of the source 
are warranted in order to find any correlated flux variations and thus 
help determine the origin of the long-term variability, which currently is
not understood.

\end{abstract}

\keywords{binaries: close --- stars: individual (2FGL~J0523.3$-$2530) --- stars: low-mass --- stars: neutron}

\section{INTRODUCTION}

Since the \textit{Fermi Gamma-ray Space Telescope} was launched in 
2008 June, the main instrument on-board---the Large Area Telescope (LAT) 
has been continuously scanning the whole sky every three hours in the energy 
range from 20 MeV to 300 GeV, discovering and monitoring $\gamma$-ray 
sources with much improved spatial resolution and sensitivity comparing to 
former $\gamma$-ray telescopes \citep{atw+09}. In 2012, using
\textit{Fermi}/LAT data of the first two-year survey, a catalog of 
1873 $\gamma$-ray sources was released by \citet{nol+12} as 
the \textit{Fermi}/LAT second source catalog (2FGL). Among the $\gamma$-ray 
sources, 575 of them are not 
associated with any known astrophysical 
objects \citep{nol+12}. For the purpose of identifying the nature of 
these unassociated sources, many follow-up studies, such as classifying 
their $\gamma$-ray characteristics \citep{ack+12}, searching for radio 
pulsars \citep{ray+12}, and observing at 
multi-wavelengths \citep{tak+13,ace+13}, have been carried out.

The source \msp\, is sufficiently bright that it was listed as
1FGL~J0523.5$-$2529 in the \fermi\  LAT First Source Catalog \citep{1fgl}. 
\textit{Swift} imaging of the field has
revealed a candidate X-ray counterpart \citep{tak+13}. While radio searches
for a pulsar have failed \citep{gui+12,pet+13}, optical imaging and
spectroscopy recently have discovered orbital modulations from 
the X-ray counterpart, with a period of 16.5 hr
(\citealt{str+14}). The source is located at a
high Galactic latitude $G_b=-29\fdg8$, and has a late-G or early-K spectral 
type secondary star, and \gr\, luminosity of 
$\sim 3.1\times 10^{33}$ erg~s$^{-1}$ (assuming source distance $d=1.1$ kpc;
\citealt{str+14}).
Based on these properties, \citet{str+14} suggested
that the source is likely a millisecond pulsar binary (MSP) 
with a 0.8~$M_\sun$ companion.
Furthermore, this binary could be another so-called ``redback" 
system, which is classified as an eclipsing MSP binary that
contains a relatively massive ($\gtrsim 0.2\,M_{\sun}$),
non-degenerate companion \citep{rob13}.
The ablation of the companion by pulsar wind from the MSP
produces matter in the binary, which would eclipse radio emission from
the pulsar at certain orbital phases.

We were intrigued by this \fermi\, source because we note
that it is located in the blazer region, along with the Crab pulsar,
in the curvature--variability plane of \fermi\ bright sources 
(see Figure~4 in \citealt{rom12}), which suggests possible variability 
from this source. It has recently been learned that the prototypical 
redback system PSR~J1023+0038 \citep{arc+09} has shown \gr\, variability 
due to its temporary accretion activity \citep{sta+14,pat+14,tak+14}. 
The newly identified redback system XSS J12270$-$4859,
which underwent a transition from an X-ray binary to a MSP binary
in 2012 Nov.--Dec. \citep{rbr14,bas+14,bog+14},
is also known to have \gr\ emission \citep{hil+11}.
Given the discovery of the orbital period of \msp\, and its properties 
listed above, we have thus carried out detailed analysis of the \fermi\ data 
for this source, aiming to study the source's \gr\, flux variations, determine
its high-energy properties, and establish the similarities to 
the redback systems PSR~J1023+0038 and XSS J12270$-$4859 in particular.
In addition, since there are two sets of archival \textit{Swift} X-ray data 
available for the source, we also conducted the X-ray data analysis.
In this paper, we report the results from the analyses.
\begin{figure*}
\centering
\includegraphics[scale=0.28]{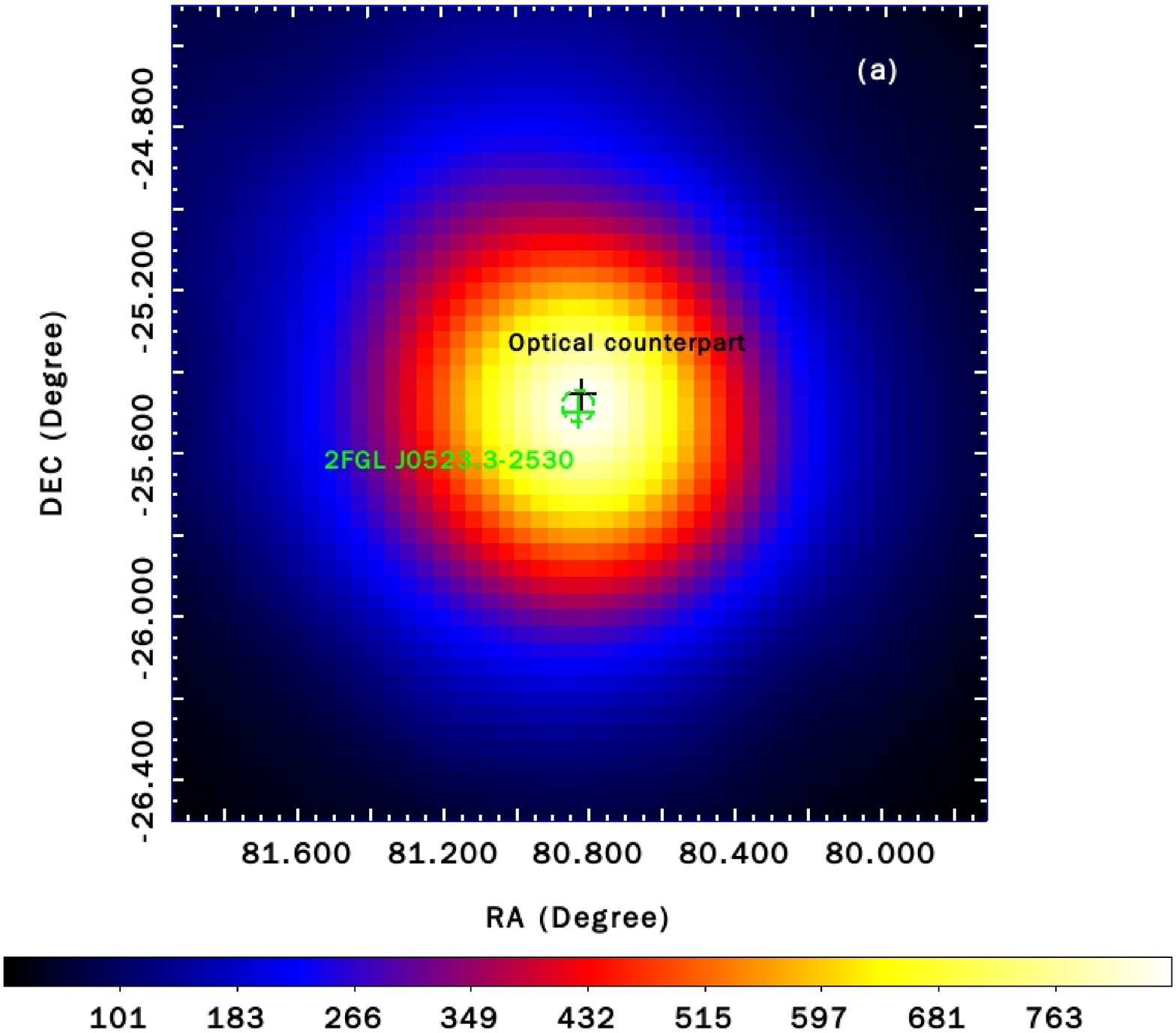}
\includegraphics[scale=0.28]{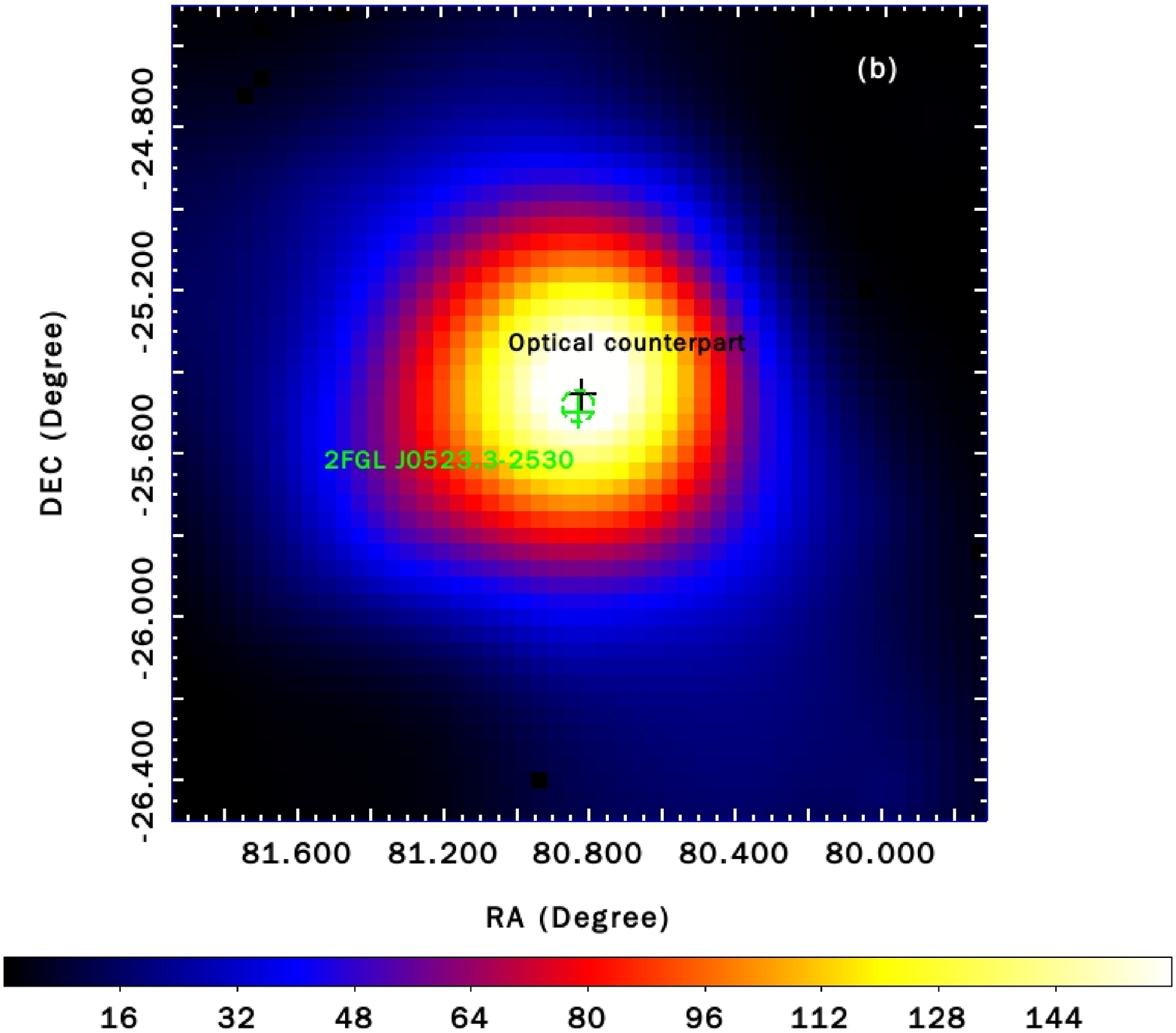}
\includegraphics[scale=0.28]{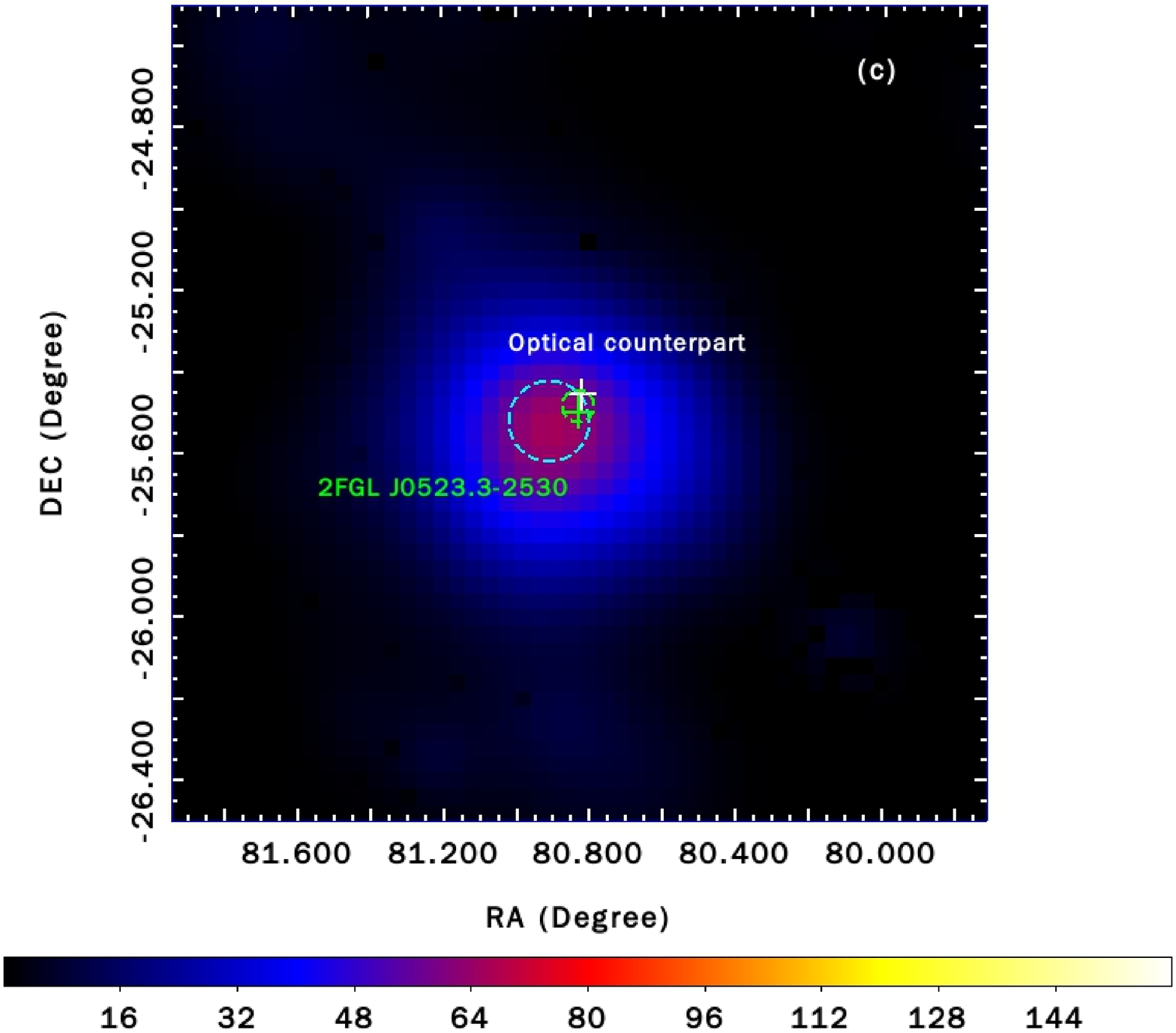}
\caption{TS maps of a $\mathrm{2^{o}\times2^{o}}$ region centered 
at 2FGL 0523.3$-$2530, with all sources in the source model 
considered and removed. Panel (a) is a 0.2--300 GeV map made 
from the whole \fermi\ data. Panels (b) and (c) 
are 1--300 GeV maps during the time interval I and II, respectively,
that are defined in \S~\ref{subsec:lv}.
The green and dark (or white in Panel (c)) crosses mark the catalog 
and optical positions, respectively, of 2FGL 0523.3$-$2530.
The green dashed circles indicate the 2$\sigma$ error circle centered 
at the best-fit position. The large circle in Panel (c) indicates
the 2$\sigma$ error circle of the source during time interval II.}
\label{fig:map}
\end{figure*}

\section{Observation}    
\label{sec:obs}

LAT is the main instrument onboard \textit{Fermi}.  It is 
a $\gamma$-ray imaging instrument which makes all-sky survey 
in an energy range from 20 MeV to 300 GeV \citep{atw+09}. 
In our analysis, we selected LAT events from the \textit{Fermi} Pass 7 
Reprocessed (P7REP) database inside a $\mathrm{20^{o}\times20^{o}}$ region 
centered at the catalog position of 2FGL J0523.3$-$2530 \citep{nol+12}. 
We kept events during the time period from 2008-08-04 15:43:36 to 
2014-04-02 01:49:57 (UTC), and rejected events below 200 MeV 
because of the relative large uncertainties of the instrument response 
function of the LAT in the low energy range. In addition we followed 
the recommendations of the LAT team to include only events with 
zenith angle less than 100\arcdeg, which prevents 
the Earth's limb contamination, and during good time intervals 
when the quality of the data was not affected by the spacecraft events. 

\section{Data Analysis and Results} 
\label{sec:ana}

\subsection{Source Identification}
\label{subsec:si}

We included all sources within 16\arcdeg\ centered at the position of 
2FGL J0523.3$-$2530 in the \textit{Fermi} 2-year catalog \citep{nol+12}
to make the source model. The spectral models of these sources 
are provided in the catalog. 
The spectral normalization parameters of the sources 
within 8 degrees from 2FGL J0523.3$-$2530 were set free, and 
all the other parameters of the sources were fixed at their catalog values. 
We also included the Galactic and extragalactic diffuse 
emission in the source model, with the spectral model 
gll\_iem\_v05.fits and the spectrum file iso\_source\_v05.txt used, 
respectively. 
The normalizations of the diffuse components were set as free parameters.
\begin{figure*}
\begin{center}
\includegraphics[scale=0.28]{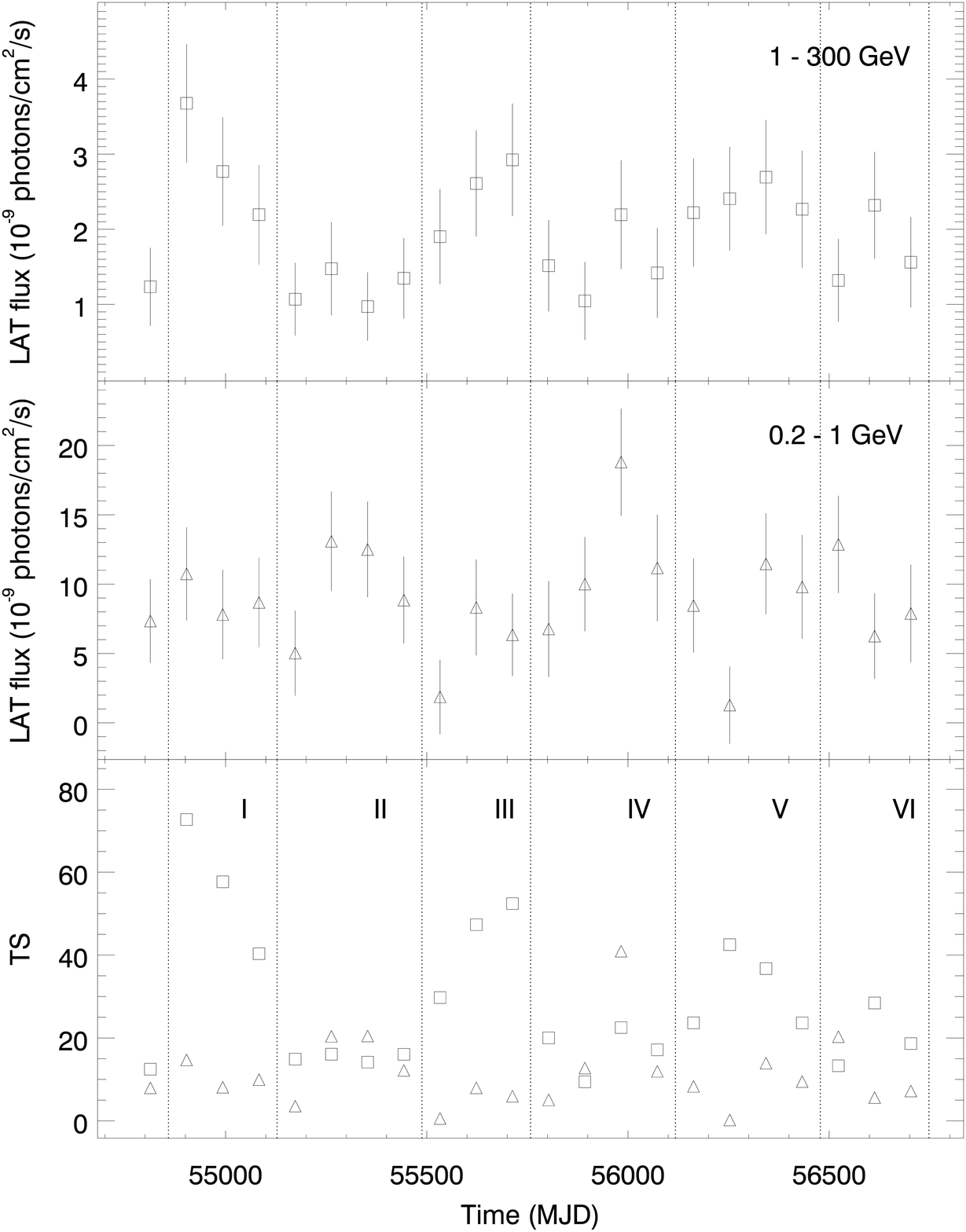}
\caption{90-day interval light curves and TS curves 
of 2FGL J0523.3$-$2530. The squares and triangles are the 1--300 GeV 
and 0.2--1.0 GeV data points, respectively. The dotted lines 
mark the six time intervals we defined.
\label{fig:lc}}
\end{center}
\end{figure*}

We performed standard binned likelihood analysis to the LAT 
$>$0.2 GeV data using the 
LAT science tools software package {\tt v9r23p5}, and extracted the 
Test Statistic (TS) map of a $2\arcdeg\times 2\arcdeg$ region centered 
at the position of 2FGL J0523.3$-$2530 
(Panel (a) of Figure~\ref{fig:map}), with all sources in the source 
model considered.
A TS value is calculated from TS $= -2\log(L_{0}/L_{1})$, where 
$L_{0}$ and $L_{1}$ are the maximum likelihood values for a model without 
and with an additional source at a specified location, respectively, and
is a measurement of the fit improvement for including the source. 
Generally the TS is approximately the square of the detection significance 
of a source \citep{1fgl}.
The $\gamma$-ray emission near the center was detected with TS$\simeq$800, 
indicating $\sim$28$\sigma$ detection significance. 
We ran \textit{gtfindsrc} in the LAT software package to find the position of 
the $\gamma$-ray emission in this region
and obtained a position of R.A.=80\fdg83, Decl.=$-$25\fdg49,
(equinox J2000.0), with 1$\sigma$ nominal uncertainty 
of 0\fdg02.
The catalog position of 2FGL J0523.3$-$2530 is R.A.=80\fdg83, 
Decl.=$-$25\fdg50 (equinox J2000.0), and the position of the optical
binary is R.A.=80\fdg8205, Decl.=$-$25\fdg4603 
(equinox J2000.0, \citealt{str+14}; the uncertainty is determined by
the USNO-B systematic accuracy of 0\farcs2, \citealt{usnob}). 
The optical position is $\sim$0\fdg03 from the best-fit position, 
but within the 2$\sigma$ error circle.

The $>$0.2 GeV emission from 2FGL J0523.3$-$2530 was analyzed 
by modeling with a simple power law and an exponentially cutoff power law 
(characteristic of pulsar \gr\,emission), respectively. 
The results are given in Table~\ref{tab:likelihood}. The source modeled with
the power-law spectrum was found to have spectral index $\Gamma=-2.17\pm 0.04$
and a TS$_{pl}$ value of $\simeq$848, while with the exponentially cutoff 
power-law 
spectrum have $\Gamma=-1.6\pm 0.1$, cutoff energy $E_{c}=4.4\pm 1.0$~GeV, and
a TS$_{exp}$ value of $\simeq$889. Therefore the cutoff 
was detected with $\sim$6$\sigma$ significance 
(estimated from $\sqrt{{\rm TS}_{exp}-{\rm TS}_{pl}}=\sqrt{41}$).
\begin{figure*}
\centering
\includegraphics[scale=0.28]{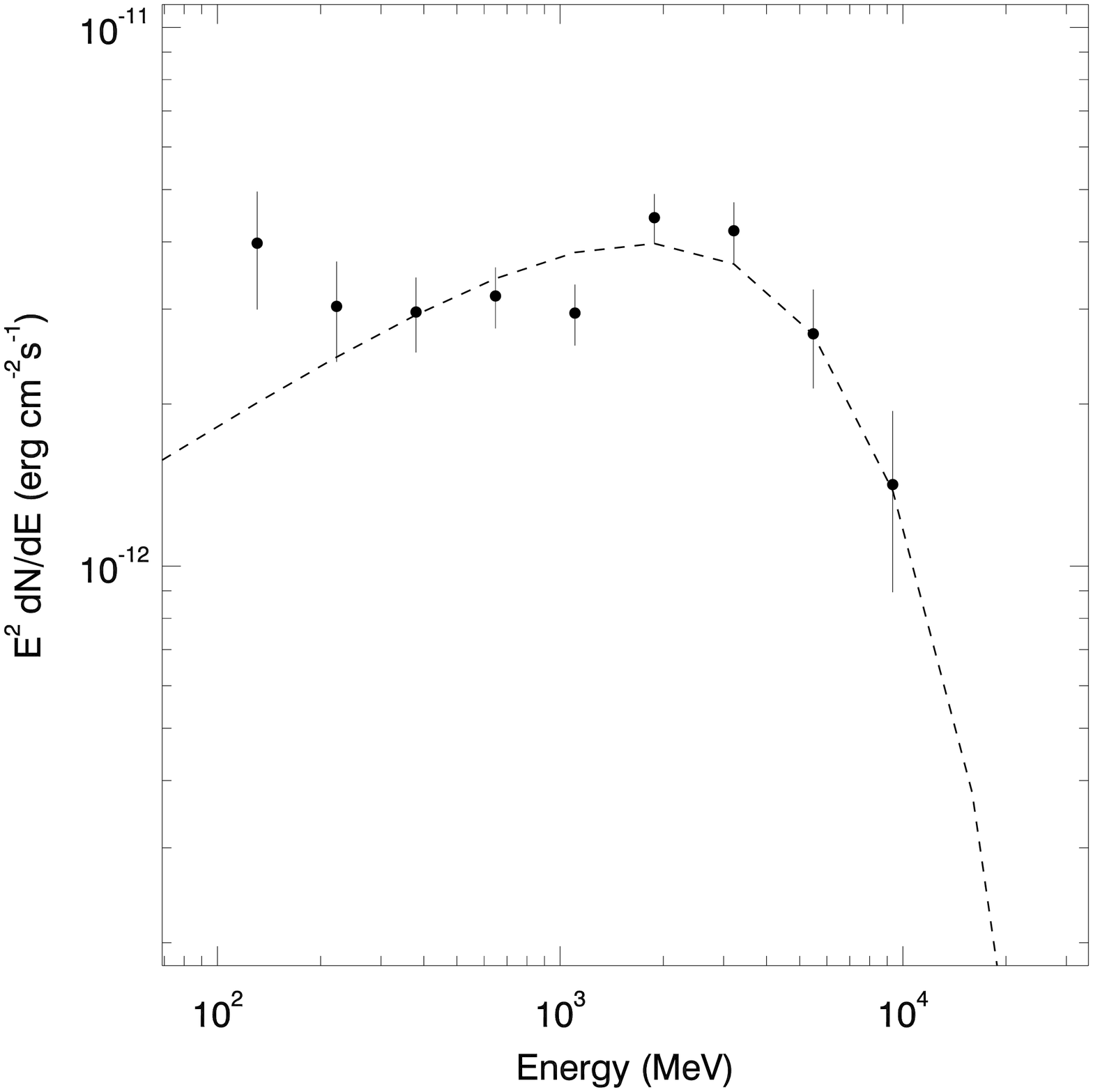}
\includegraphics[scale=0.28]{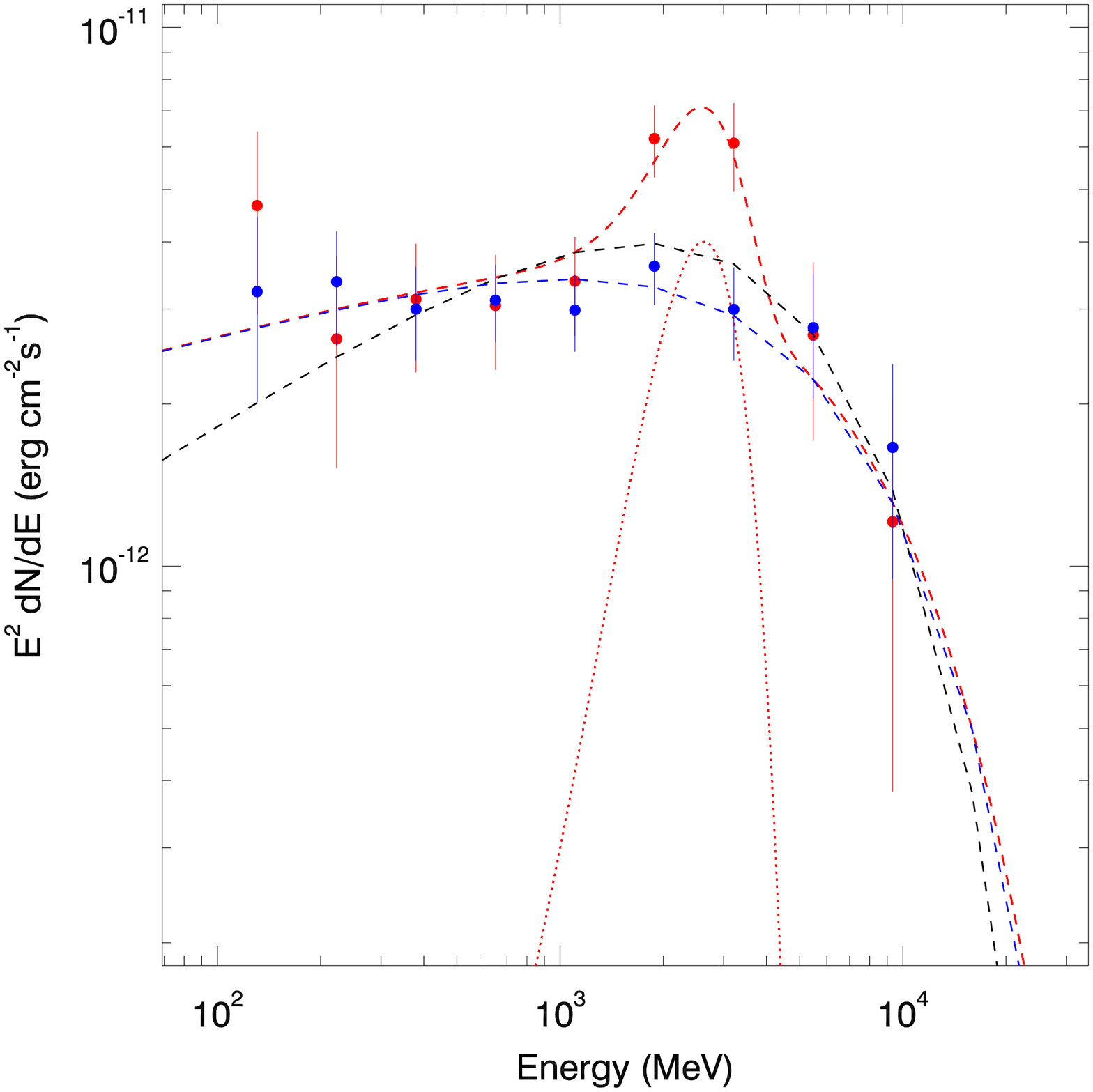}
\caption{$\gamma$-ray spectra of 2FGL J0523.3$-$2530 during the whole LAT 
observation ({\it left panel}), and during the high (red data points) and 
low (blue data points) states ({\it right panel}). 
The exponentially cutoff power law obtained from maximum likelihood analysis
for the whole data is shown as black dashed curves, and for the low state data
is plotted as the blue dashed curve. A Gaussian function (red dotted curve)
can be added to the spectrum to describe the extra component at 2-3 GeV.}
\label{fig:spectrum}
\end{figure*}

\subsection{Long-term Variability Analysis}
\label{subsec:lv}

To investigate the variability of 2FGL J0523.3$-$2530, we extracted 
different time-interval $\gamma$-ray light curves at different energy bands
($>$0.2 GeV, $>$1 GeV, or $>$2 GeV bands). The light curves were extracted 
by performing likelihood analysis to the LAT data in each time bins.
The emission of a point source with a power-law spectrum at the best-fit 
position was considered for the source.
The spectral index was fixed at the value given in 
Table~\ref{tab:likelihood}.  By comparing the obtained TS values at 
the source position,
we found that the $>$1~GeV, 90-day interval light curve shows the
most significant variations. 
In Figure~\ref{fig:lc}, we display the light curve and TS curve.
It is clear that TS value varies between 10--80, while the corresponding
flux shows a factor of 4 variation.
As a comparison, we also obtained the 0.2--1.0 GeV light curve and TS curve
and plot them in Figure~\ref{fig:lc}. The low energy TS curve generally
has low values, similar to those of the $>$1 GeV one when the latter
is in a `low' state. The flux is approximately 
10$^{-8}$ photons\,cm$^{-2}$\,s$^{-1}$, consistent with being a constant
within the uncertainties.

Based on the TS curve, we defined six time intervals. In interval II and
IV, the TS curve is flat with values of $<$20, while in interval I and III,
the TS curve has values of 30--80.
In intervals V and VI, the TS values are mostly low but with weak variations
in a range of 10--30.
To confirm the variations seen in the light curves, 
we further extracted the $>$1 GeV TS maps during the six time intervals. 
In Figure~\ref{fig:map}, we display the TS maps of intervals I and II 
as the examples for the source being bright and dim, respectively. 
The variations are real. The TS values at the source position are
$\simeq$170 and $\simeq$60 in the two maps, indicating
approximately 10$\sigma$ significance for the flux variation between the
two time intervals.

We noted that as shown in panel (c) of Figure~\ref{fig:map} 
(when the source was dim), the TS peak appears
to have an offset from the best-fit position.
We determined the source position for time interval II
and found that the position is consistent with the best-fit position
within 2$\sigma$. We further checked the TS maps when the source was dim
(time intervals IV, V, and VI), and the TS peaks all appear to have small
offsets from the best-fit position but in different directions.
In our analysis, this \fermi\ source is consistent with being a point
source, and no signs of extended emission or an additional source
were found. We concluded that the apparent offsets are probably due
to under-estimated uncertainties for the source position.


\subsection{Spectral Analysis}
\label{subsec:sa}
 
The $\gamma$-ray spectrum of 2FGL J0523.3$-$2530 was extracted by
performing maximum likelihood analysis to the LAT data 
in 15 evenly divided energy bands in logarithm from 0.1--300 GeV. 
The source was modeled with a power-law spectrum, and
the spectral index was fixed at the value we obtained above 
(Table~\ref{tab:likelihood}). 
The obtained $\gamma$-ray spectrum is shown in Figure~\ref{fig:spectrum}
and the values at each bin are given in Table~\ref{tab:spec},
in which the spectral points with TS greater than 4 were kept.
The cutoff power law model is also displayed in Figure~\ref{fig:spectrum}.
The model does not describe the low-energy data points well, as two data
points are approximately 2$\sigma$ away from the cutoff power-law model
(black dashed curve in Figure~\ref{fig:spectrum}).

To search for differences in the source's emission during the `high'
and `low' states shown in Figure~\ref{fig:lc}, and thus help understand 
the cause of the flux variation, 
we extracted $\gamma$-ray spectra of 2FGL J0523.3$-$2530 during 
the two states. We used TS=30 for defining the source's two 
states. 
The obtained spectra during the high (TS$\geq 30$) and low (TS$<$30) states
are plotted in Figure~\ref{fig:spectrum}. The flux values are given 
in Table~\ref{tab:spec}.
By comparing the two spectra, a component 
at the 2--3 GeV energy range during the high state is present. 
We therefore further modeled the low state emission with a cutoff power law,
and found $\Gamma=1.8\pm0.1$, $E_c=6.2\pm2.4$ GeV. This model is displayed
in the right panel of Figure~\ref{fig:spectrum}, showing that
it well describes the low-state spectrum and the high-state one when
excluding the 2--3 GeV data points. A Gaussian function, 
$A exp[-(E-E_0)^2/2\sigma^2]$, where 
$A=(4\pm1)\times 10^{-12}$\,erg\,cm$^{-2}$\,s$^{-1}$, $E_0=2.6\pm0.2$\,GeV,
and $\sigma=0.7\pm0.2$\,GeV, may be added to the model spectrum
to describe the extra component.  The reduced $\chi^2$, which is 
3.2 (for 6 degrees of freedom) when comparing the high-state spectrum with 
the cutoff power law, is improved to 1.0
(for 3 degrees of freedom) when the Gaussian component is included.
\begin{figure*}
\centering
\includegraphics[scale=0.28]{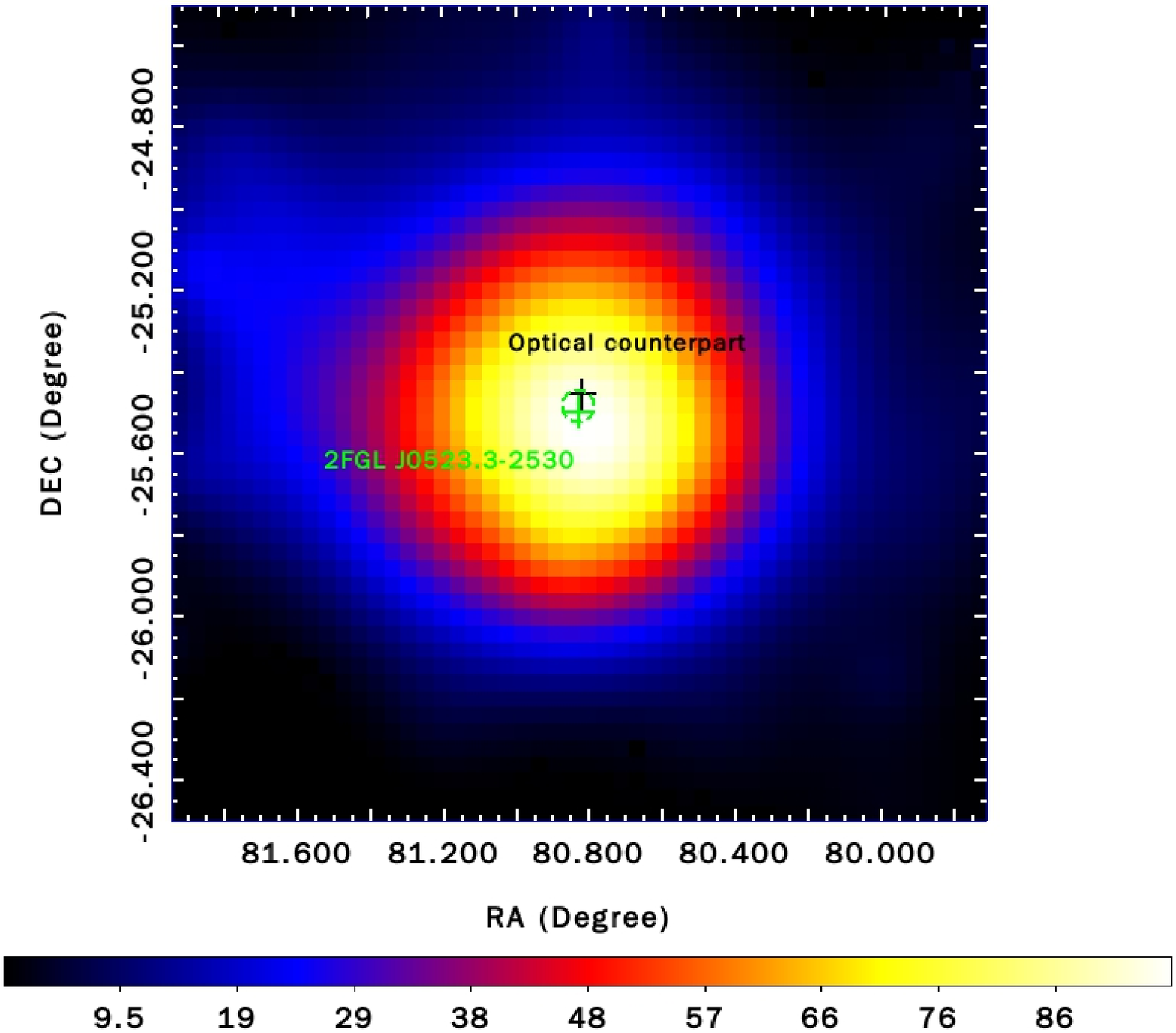}
\includegraphics[scale=0.28]{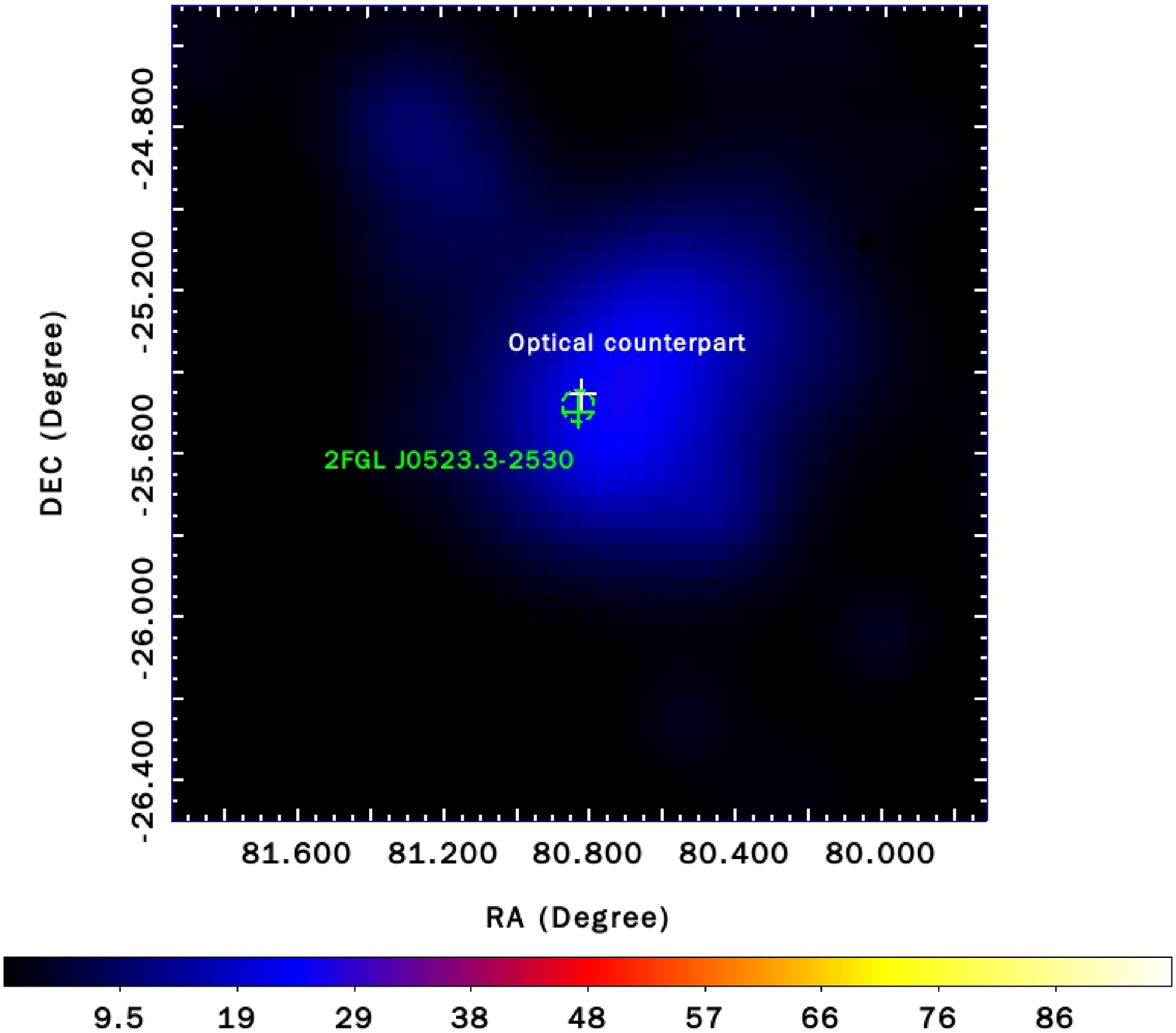}
\caption{TS maps (2--300 GeV) of a $\mathrm{2^{o}\times2^{o}}$ region 
centered at 2FGL 0523.3$-$2530 during two half orbits 
centered at the inferior conjunction ({\it left panel}) 
and superior conjunction ({\it right panel}), respectively. 
The data are from 2012-10-01 to 2014-04-02.
Symbols are the same as in Figure~\ref{fig:map}.}
\label{fig:ts-timing}
\end{figure*}

\subsection{Timing Analysis}
\label{subsec:ta}

Around the frequency 1.68195$\pm 0.00007\times 10^{-5}$ Hz, which was 
determined from optical radial velocity measurements by \citet{str+14}, 
orbital modulations were searched in \gr\,emission of the source.
The search was performed to the LAT data within 1\fdg0 
from the optical position of 2FGL J0523.3$-$2530, 
and \textit{gtpsearch} in the LAT software package
was used. The optical position was used for the barycentric correction to 
photon arrival times.
Different energy ranges of $>$0.2 GeV, $>$1 GeV, $>$2 GeV, 
and $>$3 GeV were considered in our search.
No significant signals were found for the whole data.
We also searched in the individual time periods marked by Figure~\ref{fig:lc}.
No signals were detected during the high state; in the low state
marginal signals were seen, but none of them were sufficiently convincing
as the $H$-test values for the signals were approximately 10.

Given the uncertainty of the orbital period, we considered that in
time intervals V and VI, the optical timing results are reliable (note that
the binary orbit was determined from optical observations during 
from 2013-10-01 to 2014-01-10; \citealt{str+14}) and the source
was mostly in the low state. We thus searched for periodic signals 
during a slightly longer time period from 2012-10-01 to the end of 
the LAT data we analyzed. Folding $>$2 GeV data at the optical period, 
a signal with $H$-test value of 15 was found.
Following \citet{wu+12}, we also made two TS maps over two half orbital 
phases to confirm the detection of the orbital flux variations. 
Phase I is the half of the orbit
centered at the superior conjunction (when the secondary is behind
the primary star), and Phase II is the other half centered at the
inferior conjunction.
The TS maps are shown in Figure~\ref{fig:ts-timing}.
The $>$2 GeV $\gamma$-ray emission from 2FGL J0523.3$-$2530 
during Phase II is more significantly detected than that during 
Phase I, with TS values of $\simeq$90 and $\simeq$20, respectively,
at the source position.

\begin{figure*}
\centering
\includegraphics[scale=0.28]{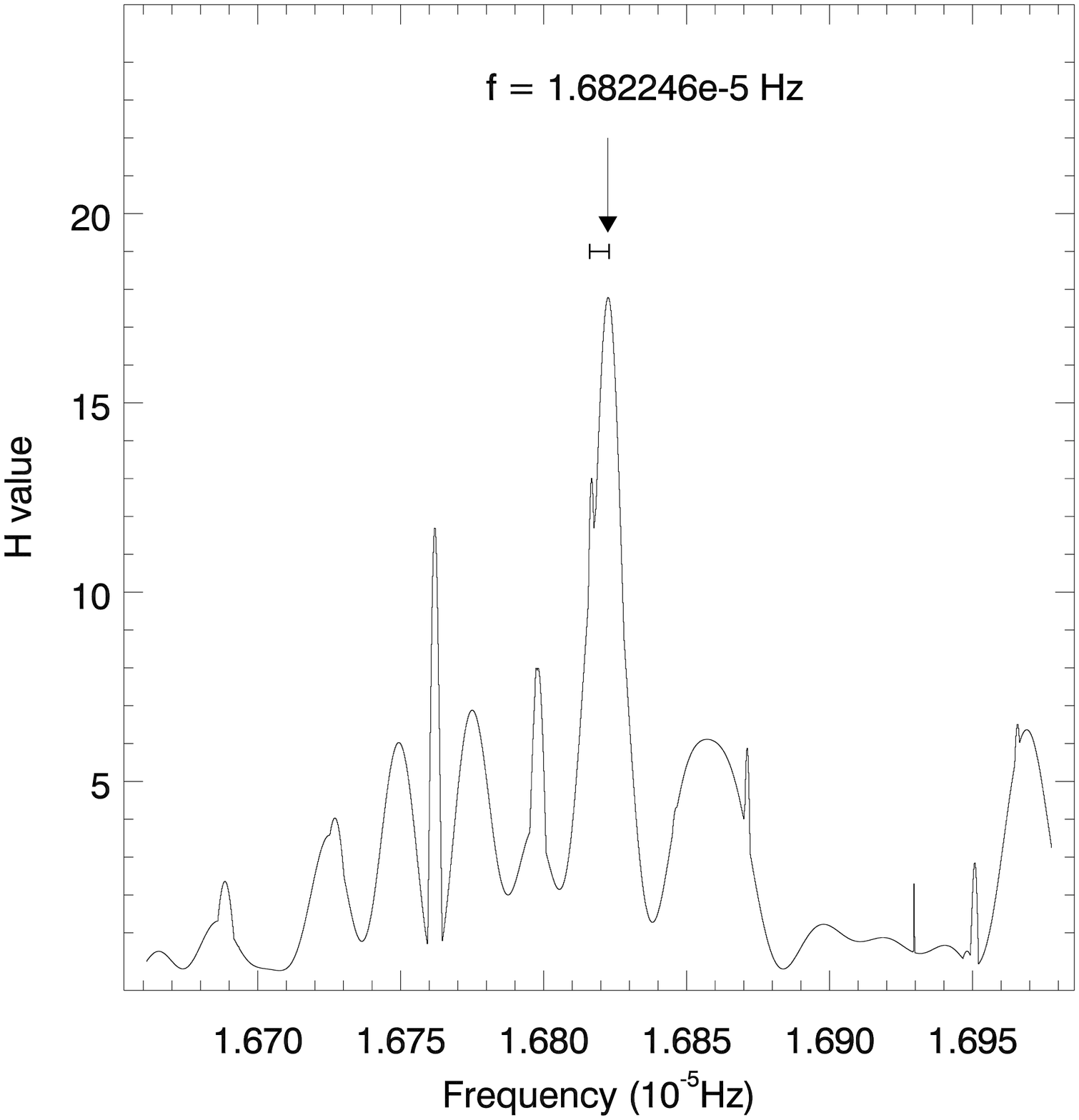}
\includegraphics[scale=0.28]{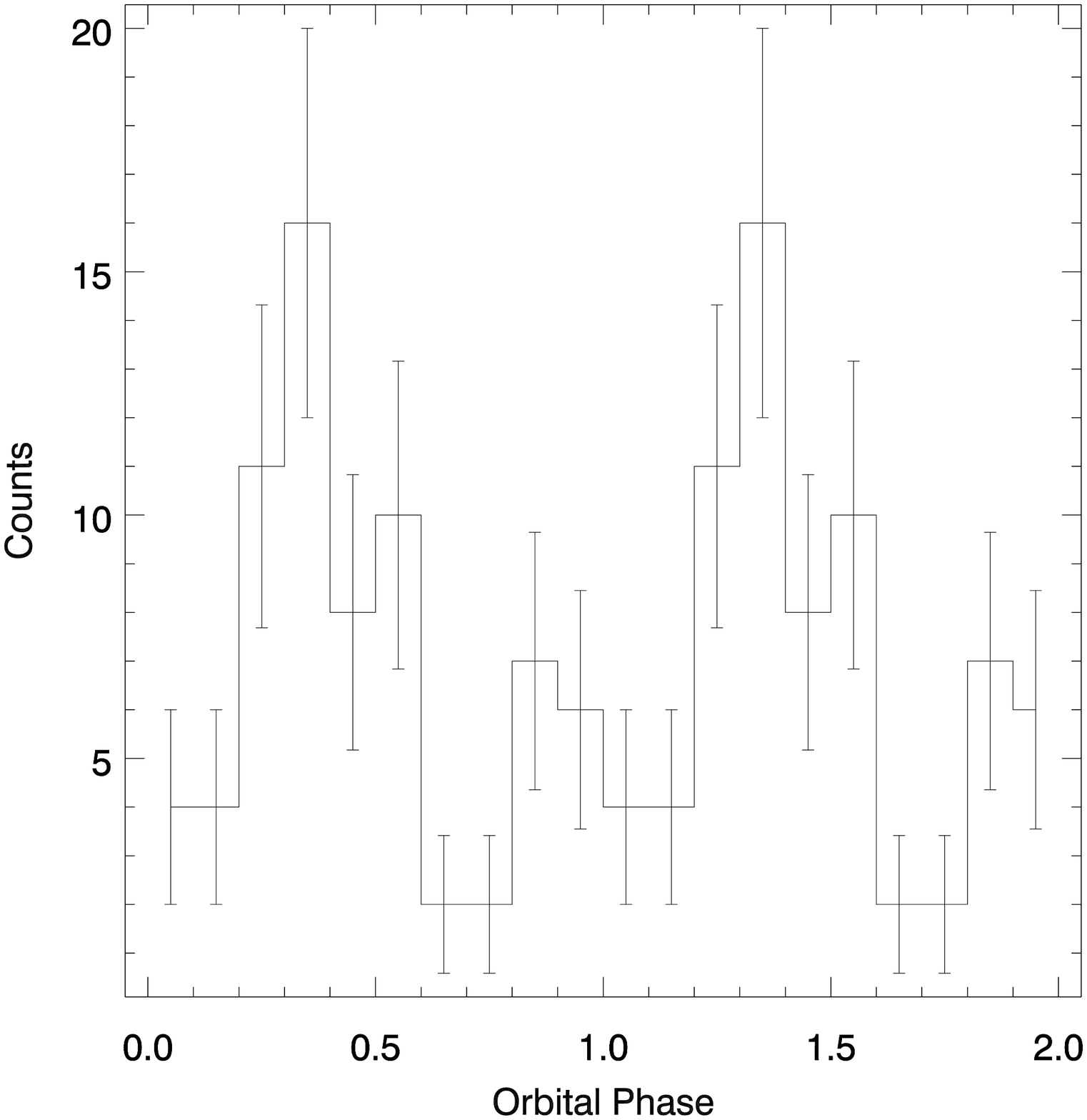}
\caption{{\it left panel:} $H$-test values at trial 
frequencies resulting from the 2012-10-01 to 2014-04-02 data. 
The frequency with the highest $H$-test value is marked by 
an arrow. 
The error bar indicates the 5$\sigma$ error range of the optical orbital
period.  {\it Right panel:} folded light curve ($>$2 GeV energy range)
at the highest $H$-test value frequency indicated in the left panel.}
\label{fig:timing}
\end{figure*}

We also searched for the periodic signals in the same energy range and
over the same time period.
A signal with $H$ test value of $\simeq$18 ($\sim$4$\sigma$ detection 
significance) at the frequency of 1.682246$\times 10^{-5}$ Hz
was found.  This frequency is within the 5$\sigma$ error range 
of the optical orbital value.  
The folded light curve is shown in Figure~\ref{fig:timing}, where
phase zero is set at the superior 
conjunction (MJD 56577.14636, given by \citealt{str+14}). 
The source was brighter during the phase of 0.25--0.55 (Phase II is
0.25--0.75). Spectra during the on-peak and off-peak phases were
obtained, but due to limited numbers of photons, the uncertainties on 
the flux data points are too large to allow any further detailed analysis.

No attempt was made to search for millisecond spin signals from
the primary star, since it is difficult and computing-intensive
to find from blind searches of \fermi\, \gr\, data, and thus far 
only one MSP has been found from blind searches \citep{ple+12}. We note
that to search for the spin signal from the putative MSP, the low-state
time periods should be considered, since emission during the time would
primarily come from the pulsar (see Section~\ref{sec:disc} below).

\subsection{\textit{Swift} X-ray Data Analysis}

The source \msp\ was observed with \emph{Swift} on 2009 Nov 12 
(ObsID: 00031535001) 
and on 2013 Sep 17 (ObsID: 00032938001) for 4.8 and 14.4\,ks,
respectively. We analyzed the photon counting mode data from the X-ray
Telescope. The data were processed by the standard pipeline,
and in both observations,
an X-ray source is clearly detected at the optical position. 
Using a standard extraction aperture of 20\,pixels (= 47\farcs1) radius, 
we obtained $14\pm4$ and $57\pm8$ background-subtracted counts 
for the first and second exposures, respectively, in 0.3--7\,keV. These
correspond to count rates of $2.9\pm0.9\times10^{-3}$ 
and $4.0\pm0.6\times10^{-3}$\,cts\,s$^{-1}$, respectively. Given
the large uncertainties, these two values are formally consistent.
The source was too faint and the two observations were too short
(comparing to the orbital period) to be searched for orbital modulation.

We extracted the source spectrum from the merged data set using the aperture
mention above. The spectral analysis was carried out in {\tt XSPEC} in the
0.3--7\,keV energy range, using the telescope response files provided by the
calibration team. For the spectral fit,
we employed the C-statistic (cstat in {\tt XSPEC}) to perform unbinned 
likelihood
analysis due to the low number of counts. We first tried an absorbed 
power-law model, but found a very small
absorption column density. This is not surprising since the source is at a
high Galactic latitude and the total Galactic column density in the direction
is only $1.7\times10^{20}$\,cm$^{-2}$ \citep{kal+05}. We therefore did
not include absorption in the final fit and obtained a photon index of
$\Gamma=1.5\pm0.2$ with a goodness-of-fit equivalent to a reduced $\chi^2$
value of 0.96. The energy flux is
$1.3\times10^{-13}$\,erg\,cm$^{-2}$\,s$^{-1}$ in 0.3--7\,keV.

\section{Discussion}
\label{sec:disc}

From our analysis of the \fermi\ data for \msp, we have found significant
\gr\ flux variations over approximately 5.5 yr \fermi\ observation time.
Spectral analysis of the data during the high and low states indicates
that emission from the source in the latter is well described by an
exponentially cutoff power law, which is typical for pulsar emission.
Comparing to the MSPs detected with \fermi\ \citep{1fpsr}, the cutoff
energy is among the highest but 
within the uncertainties (the highest value with smaller uncertainty is
$E_c=5.3\pm$1.1).
If this source is an MSP binary as suggested, the emission in the low state
is likely dominated by that from the pulsar (however, see discussion below). 
As shown in Figure~\ref{fig:lc}, in the low state, the TS values
in the energy ranges of 0.2--1 GeV and 1-300 GeV are consistently low,
not having any drastically differences as seen in the high state.

Our spectral analysis also shows that the variability mainly comes from
the presence of an extra component at 2--3 GeV.
We note that the flux changes could be rapid. For example, at the beginning
of time interval I (Figure~\ref{fig:lc}), 
the flux has a sudden increase by a factor of 4. 
To investigate the jump, we also made a 10-day interval light curve, and the 
sudden jump still exists, which suggests that the time scale
for the flux change would probably be within 10 days. This time scale
is reminiscent of PSR J1023+0038 in 2013 late June, when the pulsar binary 
was found to have an accretion disk again \citep{tak+14}. 
The \gr\ flux from the pulsar binary was also 10 times larger than 
before, a factor
of 2 times higher than what was seen in \msp. Would the flux changes be also due
to the presence of an accretion disk if the source is considered as another  
transitional pulsar binary? The non-detection of any orbital modulation
signals during the high state would support this scenario, since additional
emission from the temporary disk is suggested to be the cause of \gr\ 
brightening seen in PSR~J1023+0038 \citep{tak+14}. However, the optical 
light curve of \msp\ reported by \citet{str+14} does not support such a case.
The light curve was from data points taken from 2005 August to 2013 April,
which covers the \fermi\ observation time, but does not show any signs of 
irradiation of the companion or additional optical emission from a disk.
We also note that the two sets of X-ray data were taken 
both during the low state, which may explain the consistent faintness
of the source during the two X-ray observations.

We have detected orbital modulation in \gr\  emission of \msp, 
although only in the last 1.5 yr data of 
the total \fermi\ observation.
Unfortunately due to the limited photon counts, we were not able to
obtain any spectral information about the differences between 
the on-peak and off-peak emission.
Thus far the prototypical black window pulsar binary B1957+20, 
in which a degenerate, low-mass companion is under strong 
irradiation by pulsar wind from an MSP \citep{fst88},
is the only compact binary detected with orbital \gr\  
modulation \citep{wu+12}.  Similarly, Phase II of this binary
was found to be brighter, due to excess emission at the $>$ 2.7 GeV high 
energy range.  The excesses have been explained to be due to
inverse Compton (IC) scattering of the thermal photons from the companion
by the pulsar wind (see \citealt{wu+12} for details). The same radiation
mechanism has been considered by \citet{bed14} also for modulated \gr\  
emission from MSP binaries, although the detailed physical processes 
are different.  In any case, the modulation arises due to changes 
of the viewing angle to the intrabinary \gr\  producing region 
as the binary rotates. The high cutoff energy seen in \msp\  in the low state
could be because there is a similar extra component, arising from 
the intrabinary 
interaction. The fact that orbital modulation only appears in $>$2 GeV emission
supports this possibility. In addition, the X-ray emission
had a power-law spectrum similar to that of 
PSR J1023+0038 \citep{arc+10,bog+11} and XSS~J12270$-$4859 \citep{bog+14},
and black widow 
pulsars (e.g., \citealt{gen+14}), and the \textit{Swift} X-ray luminosity 
was 1.9$\times 10^{31}$\,erg\,s$^{-1}$
(at distance 1.1\,kpc), lower but comparable to that of PSR J1023+0038 and 
XSS~J12270$-$4859 when these two sources are at their no-disk states.
The X-ray similarity thus suggests the existence of the
intrabinary interaction as well.
We may even speculate that the yearly variability of \msp\ is
due to an enhanced intrabinary interaction of the pulsar wind 
with the companion (or outflow of the companion), resulting in
the presence of the 2--3 GeV component.
However, if this is the case, we would probably expect
easier detection of orbital modulation when the source was in
the high state, contrary to what is learned from our data analysis.


Due to the relatively large, 0\fdg02 error circle of \fermi\ sources,
it can be difficult to exclude any possible contamination from  
other sources within the error circle. For example, in the XSS~J12270$-$4859
field, a radio-jet source and a candidate galaxy cluster have been
found, which may contribute to \gr\ emission (and possible flux variations) 
detected \citep{hil+11,bog+14}. We note that 
in the SIMBAD database, none of the known nearby sources in the field 
could possibly be associated with \msp, although \citet{sch+14} 
have very recently reported the detection of a radio source 0\fdg037 away 
in their effort to search for possibly associated radio sources with \fermi\ 
objects. The majority of the high Galactic \fermi\ objects
($Gb\geq 10\arcdeg$) are associated with
Active Galactic Nuclei (AGN; \citealt{nol+12}), and they are strong
variable sources. However no evidence supports the presence of
an AGN in the field. Emission from
AGN generally has a power-law form, arising due to IC scattering by
high-energy electrons in jets from AGN \citep{abd+09}. Their
spectral features (e.g., \citealt{wil+14}) are clearly different from that 
seen in the high state
of \msp: if there is an AGN causing the variability, the flux increases
should have occurred over the whole energy range. 
Moreover, the probability of having a \gr\ emitting AGN in the
field is really low. On the basis of the AGN counts distribution study
in \citet{abd+09}, the number of AGN 
with \fermi\ \gr\ fluxes greater than 10$^{-8}$\,ph\,cm$^{-2}$\,s$^{-1}$
is 0.064 deg$^{-2}$.
For the 0\fdg06 radius (3$\sigma$) error circle region, 
the probability of having at least one such
AGN is only 6.8$\times 10^{-4}$. Given these reasons, it is not likely that the 
variability is caused by the existence of an unknown AGN.

As a summary, we have detected orbital modulation 
in $>$2 GeV \gr\ emission of \msp, which confirms the association
of this source with the optical binary. In addition, long-term, 
yearly variability from this \gr\ source has also been detected, 
and the flux increases are due to the presence
of an extra emission component at 2--3 GeV. The origin of the component
as well as the variability is not clear. In the near future, multiwavelength
studies of the source and source field should be conducted, aiming to detect
any correlated flux variations at optical/X-ray energies and help
determine the origin of the high-energy component and its variability.

\acknowledgements
We thank the anonymous referee for useful suggestions and Liang Chen
for helpful discussion about \fermi\ properties of AGN.

This research was supported by supported by Shanghai Natural Science 
Foundation for Youth (13ZR1464400), the National Natural Science Foundation
of China (11373055), and the Strategic Priority Research Program
``The Emergence of Cosmological Structures" of the Chinese Academy
of Sciences (Grant No. XDB09000000). Z.W. is a Research Fellow of the
One-Hundred-Talents project of Chinese Academy of Sciences.


\bibliographystyle{apj}

\clearpage
\begin{deluxetable}{lcccc}
\tablecaption{Binned likelihood analysis results for 2FGL J0523.3$-$2530.}
\tablewidth{0pt}
\startdata
\hline
\hline
Spectral model & Flux/$10^{-9}$ & $\Gamma$ & $E_c$ & TS \\
 &  (photon cm$^{-2}$ s$^{-1}$) &   & (GeV) &   \\
\hline
Power law & 11.5$\pm$0.7 & 2.17$\pm$0.04 & \nodata & 848 \\
Exponentially cutoff power law & 9.3$\pm$0.8 & 1.6$\pm$0.1 & 4.4$\pm$1.0 & 889 \\
Exponentially cutoff power law\tablenotemark{a} & 9.6$\pm$1.0 & 1.8$\pm$0.1 & 6.2$\pm$2.4 & 452\\
\enddata
\tablenotetext{a}{The results are from analyzing the low state data (see Section~\ref{subsec:sa}).}
\label{tab:likelihood}
\end{deluxetable}

\begin{deluxetable}{cccc}
\tablecaption{Flux measurements for 2FGL J0523.3$-$2530.}
\tablewidth{0pt}
\startdata
\hline
\hline
E & $F_{\rm low}$/10$^{-12}$ & $F_{\rm high}$/10$^{-12}$ & $F_{\rm total}$/10$^{-12}$\\
(GeV) & (erg cm$^{-2}$ s$^{-1}$) & (erg cm$^{-2}$ s$^{-1}$) & (erg cm$^{-2}$ s$^{-1}$)\\
\hline
0.13 & 3.2$\pm$1.2 & 4.7$\pm$1.7 & 4.0$\pm$1.0 \\
0.22 & 3.4$\pm$0.8 & 2.6$\pm$1.1 & 3.0$\pm$0.6 \\
0.38 & 3.0$\pm$0.6 & 3.1$\pm$0.8 & 3.0$\pm$0.5 \\
0.65 & 3.1$\pm$0.5 & 3.0$\pm$0.7 & 3.2$\pm$0.4 \\
1.10 & 3.0$\pm$0.5 & 3.4$\pm$0.7 & 2.9$\pm$0.4 \\
1.88 & 3.6$\pm$0.5 & 6.2$\pm$0.9 & 4.4$\pm$0.5 \\
3.21 & 3.0$\pm$0.6 & 6.1$\pm$1.1 & 4.2$\pm$0.5 \\
5.48 & 2.8$\pm$0.7 & 2.7$\pm$1.0 & 2.7$\pm$0.6 \\
9.34 & 1.7$\pm$0.7 & 1.2$\pm$0.8 & 1.4$\pm$0.5 \\
\enddata
\tablecomments{$F=E^2dN/dE$, obtained from the data during the low state
(column 2) and high state (column 3), and from the total data (colum 4).}
\label{tab:spec}
\end{deluxetable}

\end{document}